\documentstyle[aps,prb,floats,epsf]{revtex}
\begin{document}
\advance\textheight by 0.2in
\draft
\twocolumn[\hsize\textwidth\columnwidth\hsize\csname@twocolumnfalse%
\endcsname

\title{Monte Carlo Study of a Three--Dimensional Vortex Glass
Model with Screening}

\author{C.~Wengel and A.~P.~Young}

\address{Department of Physics, University of California, Santa
Cruz, California 95064}

\date{\today}

\maketitle

\begin{abstract}
We investigate the critical behavior of the gauge 
glass model for the vortex glass transition in three-dimensional
superconductors,
including screening of the interaction between vortices.
A Monte Carlo study of the linear resistivity and
a scaling analysis of current--voltage characteristics
indicates that screening destroys the finite--temperature
transition found earlier when screening was neglected. The 
correlation length exponent at the resulting zero temperature transition
is found to be $\nu=1.05 \pm 0.1$.
\end{abstract}
]

The question of whether type--II superconductors in the mixed state
exhibit a finite linear resistance, $\rho_{\rm lin}$, in the limit of a
vanishingly small current has been under extensive consideration
in recent years. While an applied current perpendicular to the field
generates a Lorentz force and hence 
leads to motion of flux lines producing a
voltage\cite{kim69}, it has been proposed by
Fisher\cite{fisher89} that disorder
(point defects) collectively pins the vortices, resulting in a phase
with vanishing linear resistance. Since disorder also destroys the
triangular vortex lattice predicted by mean field
theory\cite{tinkham75}, this phase is called the {\em vortex glass}.

Much of the theoretical work has been devoted to the study of a 
simplified model, called the gauge glass, which has the necessary
ingredients of randomness and frustration and is believed to be in 
the same universality class as the vortex glass. 
In two dimensions, experiments\cite{dekker92},
and simulations\cite{fty91,hyman95,bokil95}
agree that the vortex glass transition only occurs at $T_c=0$.
In $d=3$ the situation is somewhat less clear.
Several experiments on YBCO\cite{koch89} have found convincing
evidence for a finite $T_c$ phase transition, presumably to a
vortex glass. Furthermore, simulations support a transition at 
$T_c>0$, although the possibility that $T_c=0$ and the
lower critical dimension is precisely $d=3$ cannot be fully 
ruled out\cite{reger91,bokil95}.

Most of the theoretical work on the vortex glass has neglected screening
of the interaction between vortices, which arises
from coupling to the fluctuating
magnetic field. However, sufficiently close to $T_c$
the screening length, $\lambda$, and the correlation length, $\xi$,
become comparable and the
system crosses over into a region where screening is
important\cite{bokil95,ffh91}. A
recent domain wall renormalization study
of the $d=3$ gauge glass model
by one of us\cite{bokil95}
(henceforth referred to as BY)
concluded that screening apparently
destroys the finite temperature transition. For experimental systems
this effect would lead to a rounding out of the transition
close to the nominal $T_c$,
and the linear resistance would not vanish completely
but become very small. So far, experiments have not had the sensitivity
to see this effect. However, the calculations of BY were on very
small lattice sizes, $L \le 4$, and so their conclusions can only be
tentative. In this paper, we investigate the role of screening by Monte
Carlo simulations, for which much larger sizes are possible, $L \le 12$.
Our results confirm the conclusion of BY that screening destroys the
finite--$T$ transition. We are also able to obtain, for the first time,
values for the exponents at the resulting $T=0$ transition. 

The model studied here is the gauge glass with a fluctuating vector
potential
\begin{eqnarray} \label{gaugeglass}
{\cal H} & = & -J\sum_{
\langle i,j\rangle } \cos(\phi_i - \phi_j - A_{ij} - \lambda_0^{-1}
a_{ij}) \nonumber \\
& & + \frac{1}{2}\sum_{\Box}\; [\vec\nabla \times \vec a]^2,
\end{eqnarray}
where $J$ is the interaction strength, $\phi_i$ is the phase of the 
condensate on site $i$, and the sum is over all nearest neighbors
$\langle i,j\rangle$
on a simple cubic lattice. The effects of the external field
and disorder are represented by quenched vector potentials $A_{ij}$,
which we take to be uniformly distributed on the interval
$[0,2\pi]$. The fluctuating vector potentials $a_{ij}$ are integrated 
over from $-\infty$ to $\infty$, with a gauge constraint $\vec{\nabla}
\cdot \vec a = 0 $,  and $\lambda_0$ is the bare screening
length. In the last term, which describes the magnetic energy,
the sum is over all elementary
squares on the lattice and the curl is the directed sum of the vector
potentials round the square.

We consider the strong screening limit,
$\lambda_0\to0$, 
which is technically easier to simulate in the 
vortex representation. This is done by replacing the cosine in
Eq.~(\ref{gaugeglass}) with the periodic Villain function and
performing fairly standard
manipulations\cite{jose77,dasgupta81,kleinert,bokil95} to obtain 
\begin{equation}
{\cal H}_V=-\frac{1}{2}\sum_{i,j} G(i-j) \, [{\bf n}_i-{\bf b}_i] \cdot
[{\bf n}_j-{\bf b}_j].
\end{equation}
Here the ${\bf n}_i$
are vortex variables which sit on the links of the {\em dual}
lattice (which is also a simple cubic lattice),
$G(i-j)$ is
the screened lattice Green's function
\begin{equation}
G(i-j)= J {(2\pi)^2 \over L^3} \sum_{{\bf k}\neq 0}
\frac{1-\exp[{\rm i}\;{\bf k}\cdot ({\bf r}_i 
- {\bf r}_j)]}
{2\sum_{n=1}^d [1-\cos(k_n)] + \lambda_0^{-2}},
\end{equation}
(with $d=3$),
and the ${\bf b}_i$ are given by $(1/2\pi)$ times the directed sum 
of the quenched vector potential on the original lattice
surrounding the link on the dual lattice on which ${\bf b}_i$ lies.
Due to periodic boundary
conditions, we have 
the global constraints $\sum_i {\bf b}_i= \sum_i {\bf n}_i = 0$.
There are also the {\em local} constraints, 
$[\nabla\cdot {\bf n}]_i = [\nabla\cdot {\bf b}]_i = 0$, 
where the latter just follows trivially from the definition of ${\bf b}_i$
as a lattice curl.

Clearly $G(0) = 0$ and, in the limit $\lambda_0 \to 0$, $G(r) = 
J(2\pi\lambda_0)^2$ for $r\ne 0$ with corrections which are
exponentially small, {\em i.e.}\/ of order $\exp(-r/\lambda_0)$. 
Because $\sum_i ({\bf b}_i -
{\bf n}_i)= 0$ we can always add a constant to $G(r)$ for all $r$ without
affecting the results. We therefore add $-(2\pi\lambda_0)^2 J$, as a
result of which
the only interaction is on-site, and then work in units where 
$(2\pi\lambda_0)^2 J$ is unity.
The resulting Hamiltonian then has the very simple form
\begin{equation}
{\cal H}_V = \frac{1}{2}\sum_i ({\bf n}_i - {\bf b}_i)^2,
\label{H_V}
\end{equation}
which is convenient for simulations. 
Note, however, that
${\cal H}_V$ is not trivial because the local constraint
$[\nabla\cdot {\bf n}]_i = 0$ effectively generates interactions
between the ${\bf n}_i$.
Note also that without disorder
({\em i.e.} all ${\bf b}_i=0$) this model is just a dual representation of
the $XY$--model\cite{dasgupta81,kleinert} (with the Villain potential),
in which the temperature scale has been inverted.

We simulate the Hamiltonian in Eq.~(\ref{H_V}) on simple cubic
lattices with $N=L^3$ sites where $4 \le L \le 12$. Periodic boundary
conditions are imposed. 
We start with configurations with all ${\bf n}_i = 0$, which clearly
satisfies the constraints, and a Monte Carlo move consists
of trying to create a loop of 
four vortices around a square. This trial state is accepted  with probability 
$1/(1+\exp(\beta \Delta E))$,
where $\Delta E$ is the change of energy and $\beta=1/T$.
Each time a loop is 
formed it generates a voltage
$\Delta Q = \pm 1$ perpendicular to it's plane, the sign
depending on the orientation of the loop. This leads 
to a net voltage\cite{hyman95}
\begin{equation}
V(t)=\frac{h}{2e} I^V(t)
\quad\mbox{with}\quad
I^V(t)=\frac{1}{L\Delta t}  \Delta Q(t) ,
\label{voltage}
\end{equation}
where $I^V$ is the vortex--current
and $t$ denotes Monte Carlo ``time'' incremented by
$\Delta t$ for each attempted Monte Carlo move.
We will work in units where $h/(2e)=1$, and we set $\Delta t=1/3N$
so that an attempt is made to create or destroy one vortex loop
per square in each direction, on average, per unit time.

The linear resistivity
can be calculated from the voltage fluctuations via the 
Kubo formula\cite{young94}
\begin{equation}
\rho_{\rm lin}=\frac{1}{2T}\sum_{t=-\infty}^{\infty} \Delta t\; 
\langle V(t)V(0)\rangle.
\end{equation}
Near a second order phase transition the linear resistivity obeys the
scaling law\cite{ffh91}
\begin{equation}
\rho_{\rm lin}(T,L)=L^{-(2-d+z)} \tilde{\rho}(L^{1/\nu}(T-T_c)),
\label{rho_lin_scale}
\end{equation}
where $\tilde{\rho}$ is a scaling function and $z$ is the dynamical 
exponent. At the critical temperature, $\tilde{\rho}$ becomes a 
constant and therefore $\rho_{\rm lin}(T_c,L) \sim L^{-(2-d+z)}$.
If we plot the ratio
$\ln[\rho_{\rm lin}(L)/\rho_{\rm lin}(L^{\prime})]/\ln[L/L^{\prime}]$
against $T$,
then at the point ($T_c$, $d-2-z$) all curves for different pairs 
$(L,L^{\prime})$ should
intersect and one can read off the values of $T_c$ and $z$. We will refer to
this kind of data plot as the ``intersection method''.

In addition to $\rho_{\rm lin}$, we also measure
the voltage generated by
a finite external current. In real superconductors, transport currents
generate a non-uniform magnetic field because of Amp\`ere's law,
$\vec \nabla \times {\bf B} = {\bf J}$. It is inconvenient to simulate a
non--uniform system, so instead we
effectively assume that the current is the same everywhere so
{\em each} vortex feels a Lorentz force ${\bf n}_i \times {\bf J}$.
The scaling behavior of the response to such a perturbation should be
the same as that derived earlier for response to an actual
transport current\cite{ffh91}. We can therefore use this approach to
determine critical exponents, which is our objective. 
The Lorentz force biases the moves and sets up a net flow of vortices
perpendicular to the current, whose time average gives the voltage
according to\cite{hyman95} Eq.~(\ref{voltage}).

To analyze our data we need to understand the scaling behavior
of the $I$-$V$--curves near a second order phase transition. We 
quickly review the scaling theory\cite{ffh91} 
assuming $T_c=0$ so the correlation length 
diverges as $\xi\sim T^{-\nu}$.
The vector potential enters the Hamiltonian in Eq. (\ref{gaugeglass})
in the dimensionless form $A_{ij}=\int_i^j {\bf A}({\bf r})\cdot d{\bf r}$
so ${\bf A}$ scales as $1/\xi$. The electric field is given
by ${\bf E}=-\partial_t {\bf A}$ and so scales as $1/(\xi\tau)$ where
$\tau$ is the relaxation time. ${\bf J}\cdot {\bf E}$ is 
the energy dissipated per unit volume and unit time and 
scales as\cite{footnote1} $T/(\xi^d \tau)$. Hence $J$ scales 
like $T/\xi^{d-1}$. Combining these results leads to
\begin{equation}
T\frac{E}{J}\frac{\tau}{\xi^{d-2}}=g\left(\frac{J\xi^{d-1}}{T}\right),
\end{equation}
where $g$ is a scaling function. In three dimension this becomes
\begin{equation}
T^{1+\nu} \frac{E}{J} \tau =g\left(\frac{J}{T^{1+2\nu}}\right).
\end{equation}
From this equation we can see that the current scale, $J_{NL}$,
at which nonlinear
behavior sets in varies with $T$ as $J_{\rm NL}\sim T^{1+2\nu}$.
Since the linear resistivity is defined by
\begin{equation}
\rho_{\rm lin}=\lim_{J\to 0}\frac{E}{J},
\end{equation}
and $g(0)$ can be taken to be unity, we can write
\begin{equation}
\label{scale_iv}
\frac{E}{J\rho_{\rm lin}}=g\left(\frac{J}{T^{1+2\nu}}\right).
\end{equation}
Furthermore, we expect that near the $T=0$ transition, long time
dynamics will be governed by activation over barriers. Hence we expect
\begin{equation}
T^{1+\nu}\rho_{\rm lin}=\frac{1}{\tau}=A\exp(-\Delta E(T)/T),
\end{equation}
where $\Delta E$ is the typical barrier that a vortex has to 
cross to move a distance $\xi$.	
One can define a  barrier height exponent $\psi$ by
$\Delta E\sim \xi^{\psi}\sim T^{-\psi\nu}$ in terms of which 
\begin{equation}
T^{1+\nu}\rho_{\rm lin}=A\exp(-C/T^{1+\psi\nu}).
\label{arrhenius}
\end{equation}
We are able to obtain a rough estimate for $\psi$ from 
our data of the linear resistivity.

In a finite system, the $I$-$V$ characteristics will also
depend on the ratio $L/\xi$. One can generalize
the scaling function, Eq.~(\ref{scale_iv}), to account for finite
size effects as follows:
\begin{equation}
\frac{E}{J\rho_{\rm lin}} = \tilde{g}\left( \frac{J}{T^{1+2\nu}},L^{1/\nu}T
\right).
\label{iv-scale}
\end{equation}
Now we are left with a rather complicated scaling function
since it depends on two variables. We estimated $\nu$ by determining the
current where $E / (J\rho_{\rm lin}) = 2$, at which point non-linear effects
start to become significant. Denoting these values of $J$ by $J_{\rm NL}$,
then, from Eq.~(\ref{iv-scale}), it follows that
$J_{\rm NL}/T^{1+2\nu}$ is a function of $L^{1/\nu}T$. Collapsing the
data in the appropriate plot then gives an estimate of $\nu$. We then
collected data for sizes and temperatures such that $L^{1/\nu}T$ is
constant so the scaling function in Eq.~(\ref{iv-scale}) then
depends only on one variable.


\begin{figure}[t]
\epsfxsize=\columnwidth\epsfbox{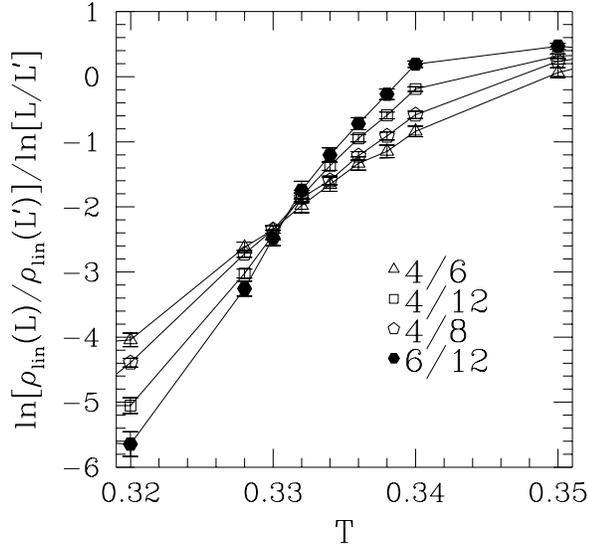}
\caption{Plot of 
$\ln[\rho_{\rm lin}(L)/\rho_{\rm lin}(L^{\prime})]/\ln[L/L^{\prime}]$
versus $T$ for 
the pure system, whose Hamiltonian is given by Eq.~(\protect\ref{H_V}) with
${\bf b}_i = 0$. This is a dual representation of the pure $XY$ model
(with the Villain potential) with inversion of the temperature scale.
The curves intersect at $T_c=0.331 \pm 0.002$. At the intersection
point,
the $y$--value is approximately $-2$ corresponding to $z \simeq 3$ according to
Eq.~(\protect\ref{rho_lin_scale}).}
\label{interpure}
\end{figure}

Let us now move to the analysis of our data. 
In Fig.~\ref{interpure} we show data of the linear resistivity 
for the pure case plotted
according to the intersection method. All curves intersect at 
about $T=0.331 \pm 0.002 $ which is in excellent agreement with the well 
established value\cite{dasgupta81,jose77}
of $T_c=0.33$.
The
$y$--axis value at the intersection point is $1-z$ according to
Eq.~(\ref{rho_lin_scale}), 
from which we estimate
$z\simeq 3 $. This is larger than one would expect for a model with
relaxation dynamics where usually $z \simeq 2$. Perhaps corrections to
finite size scaling make the estimates of $T_c$ and $z$ slightly
inaccurate, with the effect on $z$ being more pronounced since the
slopes of the curves in Fig.~\ref{interpure} are quite steep.
Nonetheless, we are interested in whether or not a transition occurs
rather than precise values of exponents, and
the method seems to be a reliable in predicting this, at least for the
pure case.

\begin{figure}[t]
\epsfxsize=\columnwidth\epsfbox{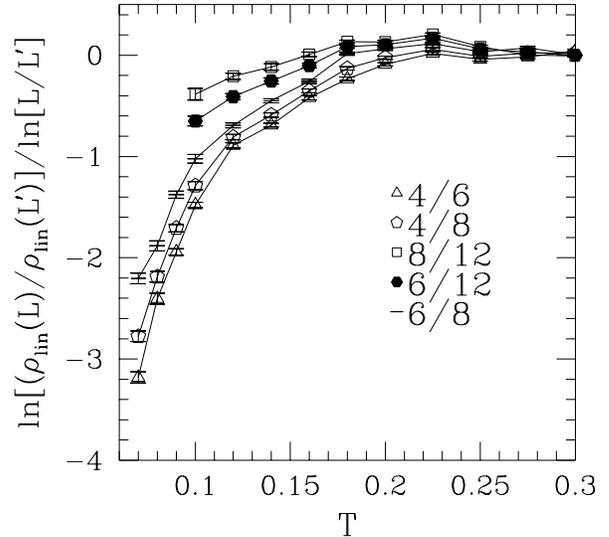}
\caption{Plot of
$\ln[\rho_{\rm lin}(L) /
\rho_{\rm lin}(L^{\prime})]/\ln[L/L^{\prime}]$ against
$T$ for the gauge glass model whose Hamiltonian is given by
Eq.~(\protect\ref{H_V}).
In contrast to the data for the pure system in
Fig.~\protect\ref{interpure}, there is no 
intersection over the entire temperature range, indicating the absence
of a phase transition.}
\label{interdis}
\end{figure}

For comparison, Fig.~\ref{interdis} shows the intersection
method applied to the gauge glass model. As one can see, there is no apparent
intersection
over the entire temperature range that we have been able to simulate,
{\em i.e.}
down to $T=0.1$ for $L=12$ and down to $T=0.07$ for
$L \le 8$. This rules out a transition down
to about $1/5$ of the critical temperature of the pure system, and
strongly
suggests that the phase transition is at zero temperature.

\begin{figure}[t]
\epsfxsize=\columnwidth\epsfbox{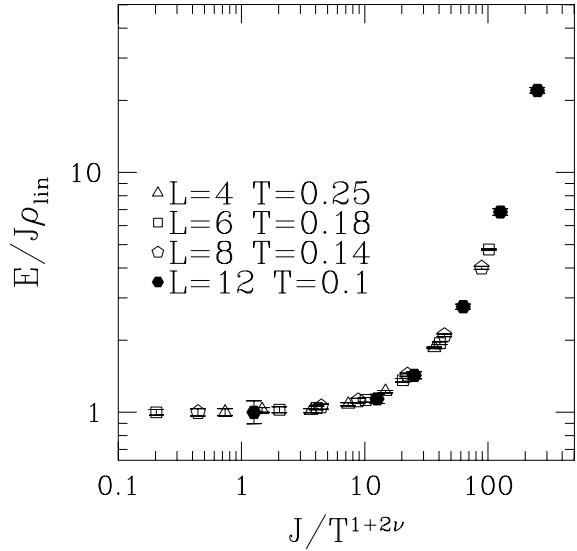}
\caption{Scaling plot of the $I$-$V$--characteristics with $T_c=0$
and $\nu=1.05 $, according to Eq.~(\protect\ref{iv-scale}),
choosing sizes and temperatures such that
$L^{1/\nu}T$ is roughly constant.}
\label{ivscale}
\end{figure}

Fig.~\ref{ivscale} shows a scaling plot of the nonlinear
$I$-$V$ characteristics according to Eq.~(\ref{iv-scale}).
From the scaling of
$J_{\rm NL}$ we estimated $\nu$ to be approximately 1.1,
and then chose temperatures and sizes for the data in Fig.~\ref{ivscale}
such that
$L^{1/\nu}T$, and hence the second argument in Eq.~(\ref{iv-scale}),
remained roughly constant.
The data is seen to scale very well.
Combining this data together with
scaling plots
for other sizes and temperatures we obtained the overall
estimate $\nu=1.05 \pm 0.1$.
We also tried to scale our data with an appropriate scaling
function for finite $T_c$ and found that scaling works
satisfactorily with $T_c\simeq 0.04$ and $\nu\simeq 1.1$.
Therefore, we conclude that the transition is very likely to
occur at $T_c=0$, although a finite but extremely small $T_c$
cannot be completely ruled out.

\begin{figure}[t]
\epsfxsize=\columnwidth\epsfbox{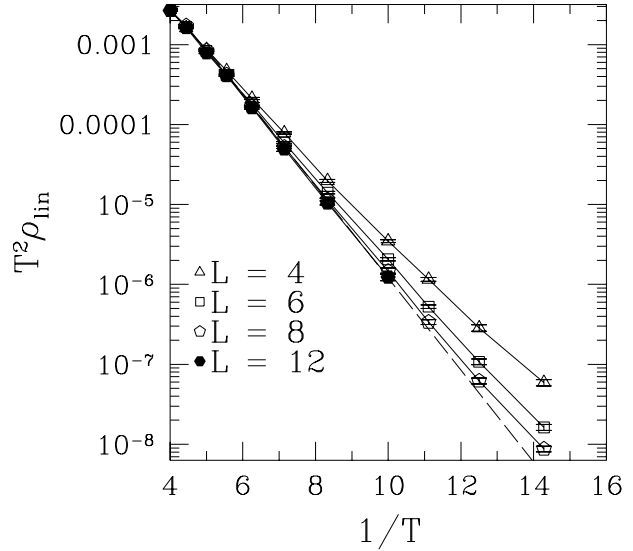}
\caption{Plot of $T^2 \rho_{\rm lin}$ a
logarithmic scale versus $1/T$.
The dashed line corresponds to an Arrhenius law, which, according to
Eq.~(\protect\ref{arrhenius}), corresponds to a barrier exponent,
$\psi$, equal to zero.
The data for the largest size, $L=12$, is well approximated by this.}
\label{rhodis}
\end{figure}

Finally, Fig.~\ref{rhodis} shows data for $T^2 \rho_{\rm lin}$ 
versus $1/T$ in a log--linear plot according to
Eq.~(\ref{arrhenius}).
The data for $L=12$ follows
almost a straight line, {\em i.e.} shows close to Arrhenius behavior, which
corresponds to a barrier exponent $\psi = 0$
according to Eq.~(\ref{arrhenius}).
It is therefore possible
that the barriers only increase logarithmically as the temperature 
approaches zero. However, such behavior is difficult to observe 
in finite size systems over a modest range of temperatures.
It is also possible that we are only measuring an effective
exponent and that the true $\psi$ is larger than zero.


In summary, we have performed a Monte Carlo study of the $d=3$ gauge
glass model in the vortex 
representation with strong screening. We have analyzed the
linear resistivity 
and $I$-$V$ characteristics
by means of finite size scaling. There appears to be no transition
of this model into a vortex glass state at finite temperature
--- consistent with the findings of BY ---
and the 
correlation length exponent at the resulting zero temperature transition
is found to be $\nu=1.05 \pm 0.1$. 
However, the possibility that $T_c$ is small but non--zero
cannot be completely ruled out
since we are not 
able to equilibrate a range of system sizes below about 1/5 of the
transition temperature of the pure system.
Assuming $T_c = 0$, the presumed finite temperature
vortex glass transition in the three--dimensional gauge glass 
is destroyed by screening effects.
This means that the linear resistance in 
experimental systems would not strictly vanish, though
it would become extremely small for $T$ near the apparent $T_c$.
It would be interesting to
look for this effect experimentally.

We would like to thank Hemant Bokil for illuminating discussions. The
work of APY was supported by NSF grant DMR 94-11964.
The work of CW was supported by the German Academic Exchange Service
(Doktorandenstipendium HSP II/AUFE).


\begin{thebibliography}{99}
\bibitem{footnote1}
  Note that it is important to keep the factor of $T$ for a 
  zero temperature transition.
\end{thebibliography}
\end{document}